\documentclass{elsart}
\usepackage{natbib,psfig}
\begin{document}

\newcommand{\ndetz}{11}
\newcommand{\nobsz}{23}
\newcommand{\ndetnoz}{4}
\newcommand{\nobsnoz}{10}
\newcommand{\ndettot}{15}
\newcommand{\nobstot}{33}
\newcommand{\ndetzAR}{15}
\newcommand{\tur}{4C~41.17}
\newcommand{\squ}{4C~60.07}
\newcommand{\hor}{B2~0902+34}
\newcommand{\boa}{MRC~1243+036}
\newcommand{\fis}{MRC~1138$-$262}
\newcommand{\arch}{A01}
\newcommand{\eg}{{e.g.,}}
\newcommand{\lya}{\ifmmode {\rm Ly\alpha}\else{\rm Ly$\alpha$}\fi}
\newcommand{\hzrgs}{HzRGs}
\newcommand{\hzrg}{HzRG}
\newcommand{\lss}{LSS}
\newcommand{\LFIR}{\ifmmode {\rm \,L_{FIR}}\else ${\rm \,L_{FIR}}$\fi}
\newcommand{\LUV}{\ifmmode {\rm \,L_{UV}}\else ${\rm \,L_{UV}}$\fi}
\newcommand{\Llya}{\ifmmode {\rm \,L_{Ly\alpha}}\else ${\rm \,L_{Ly\alpha}}$\fi}
\newcommand{\OmM}{\ifmmode {\Omega_{\rm M}}\else $\Omega_{\rm M}$\fi}
\newcommand{\OmL}{\ifmmode {\Omega_{\Lambda}}\else $\Omega_{\Lambda}$\fi}
\newcommand{\ph}{\ifmmode {h_{65}^{-1}}\else $h_{65}^{-1}$\fi}
\newcommand{\psqh}{\ifmmode {h_{65}^{-2}}\else $h_{65}^{-2}$\fi}
\newcommand{\arcsec}{\ifmmode {\,''} \else {$\,''$}\fi}
\newcommand{\degree}{\ifmmode {^{\,\circ}} \else {$^{\,\circ}$}\fi}
\newcommand{\Lsun}{\ifmmode {\rm\,L_\odot}\else ${\rm\,L_\odot}$\fi}
\newcommand{\Msun}{\ifmmode {\rm\,M_\odot} \else ${\rm\,M_\odot}$\fi}
\newcommand{\Zsun}{\ifmmode {\rm\,Z_\odot} \else ${\rm\,Z_\odot}$\fi}
\newcommand{\kmps}{\ifmmode {\rm\,km~s^{-1}} \else ${\rm\,km\,s^{-1}}$\fi}
\newcommand{\kpc}{{\rm\,kpc}} 
\newcommand{\ergps}{\ifmmode {\rm\,erg\,s^{-1}} \else {${\rm\,erg\,s^{-1}}$}\fi}
\newcommand{\ergpspcm}{\ifmmode {\rm\,erg\,s^{-1}\,cm^{-2}} \else {${\rm\,erg\,s^{-1}\,cm^{-2}}$}\fi}
\newcommand{\micron}{\ifmmode {\mu {\mathrm m}} \else {$\mu {\mathrm m}$}\fi}
\newcommand{\surfbr}{\ifmmode {\rm\,erg\,s^{-1}\,cm^{-2}\,arcsec^{-2}} \else {${\rm\,erg\
,s^{-1}\,cm^{-2}\,arcsec^{-2}}$}\fi}
\newcommand{\Msunpyr}{\ifmmode {\rm\,M_\odot\,yr^{-1}} \else {${\rm\,M_\odot\,yr^{-1}}$}\fi}
\newcommand{\pyr}{\ifmmode {\rm\,yr^{-1}} \else {${\rm\,yr^{-1}}$}\fi}
\newcommand{\psec}{\ifmmode {\rm\,s^{-1}} \else {${\rm\,s^{-1}}$}\fi}
\newcommand{\aap}{{A\&A}}
\newcommand{\aaps}{{A\&A}}
\newcommand{\aj}{{AJ}}
\newcommand{\apj}{{ApJ}}
\newcommand{\apjl}{{ApJ}}
\newcommand{\araa}{{ARA\&A}}
\newcommand{\mnras}{{MNRAS}}
\newcommand{\nat}{{Nature}}
\newcommand{\procspie}{{Proc. SPIE}}



\runauthor{Michiel Reuland}
\begin{frontmatter}
\title{SCUBA Observations of High Redshift Radio Galaxies}
\author[IGPP,Leiden,Davis]{Michiel Reuland}
\author[Leiden]{Huub R\"ottgering}
\author[IGPP]{Wil van Breugel}

\address[IGPP]{IGPP/LLNL, L-413, P.O. Box 808, Livermore, CA 94551, U.S.A.}
\address[Leiden]{Leiden Observatory, P.O. Box 9513, 2300 RA Leiden, The Netherlands}
\address[Davis]{Department of Physics, UC Davis, 1 Shields Avenue, Davis, CA 95616, U.S.A.}

\begin{abstract}
High redshift radio galaxies (\hzrgs) are key targets for studies of
the formation and evolution of massive galaxies.  The role of dust in
these processes is uncertain. We have therefore observed the dust
continuum emission from a sample of $z > 3$ radio galaxies with the
SCUBA bolometer array.  We confirm and strengthen the result found by
\citet{Archibald01mnras}, that \hzrgs\ are massive starforming systems
and that submillimeter detection rate appears to be primarily a strong
function of redshift.  We also observed \hzrg-candidates which have
sofar eluded spectroscopic redshift determination. Four of these have
been detected, and provide evidence that they may be extremely
obscured radio galaxies, possibly in an early stage of their
evolution.
\end{abstract}
\end{frontmatter}

\section{Introduction}
The discovery of a tight relation between the masses of the stellar
bulges of galaxies and the masses of their central black holes
\citep{Magorrian98aj} has reemphasized the importance of high redshift
active galactic nuclei (AGNs) as probes of galaxy formation. They are
therefore the target of many studies of starformation in their host
galaxies.  Since a large proportion of the optical and UV emission
from a starburst is absorbed and reprocessed by dust to restframe
far-IR wavelengths, a good way to investigate the roles of dust and
star formation in these distant galaxies is to measure their
\mbox{(sub-)millimeter} continuum emission.  Another advantage of
observing at those wavelengths is that dimming due to cosmological
distance is compensated for as the peak in the dust emission is
redshifted into the observed passband
\citep[\eg][]{BlainLongair93mnras}

The formation of massive elliptical galaxies may be best studied by
tracing the evolution of powerful radio galaxies.  They require the
presence of an actively-accreting super massive black hole
\citep[$\sim 10^{9}\,\Msun$;][]{Lacy01apj}, are very extended and
luminous at many wavelenghts allowing detailed studies, and their
selection at radio wavelengths is not affected by the presence of
dust.

\citet[][]{Archibald01mnras} have conducted a submillimeter survey of
47 radio galaxies over an interval of $0.7 < z < 4.4$. They found a
substantial increase in 850\,\micron\ detection rate with redshift and
that the average 850\,\micron\ luminosity $L_{850}$ rises at a rate
$(1 + z )^{3-4}$ out to $z \simeq 4$.  While it is true that a buried
AGN could be responsible for heating the dust, there are strong
indications that starbursts (SFR $\sim 1000 \Msunpyr$) are the dominating mechanism: (i) there is
no strong correlation between radio power and submillimeter luminosity
\citep{Archibald01mnras,Reuland03prepb}, (ii) for all hyperluminous
infrared galaxies the starburst outshines the AGN at rest-frame
wavelengths longwards of 50\,\micron\
\citep{RowanRobinson00mnras,Farrah02mnras}, and (iii) the extent (up
to 30\,kpc) of the CO and dust-emission found in 3 \hzrgs\
\citep{Papadopoulos00apj,DeBreuck02prep} is not readily explained by a
central process.

Until recently only few HzRGs at $z > 3$ were known, hampering
statistical studies. However, using radio and near-IR selection we
have significantly increased the sample of $z > 3$ \hzrgs\ at Keck
\citep[\eg][]{DeBreuck01aj}. Here we present submillimeter
observations of these. Additionally, we have observed an intriguing
class of \hzrg-candidates for which no spectroscopic redshift could be
determined.  One of these candidates had been studied before in detail
\citep[WN~J0305+3525;][]{Reuland03apj} and was found to be very
submillimeter luminous.

\section{Sample Selection and Observations}

We have selected all known \hzrgs\ with redshifts $z > 3$ without
prior SCUBA observations.  We aimed to observe a significant sample of
\hzrgs\ to complement the observations of \citet{Archibald01mnras} and
obtain better statistical info at the highest redshifts.  The sample
is the result of an ongoing effort by our group and others
\citep[\eg][]{DeBreuck01aj} to find distant radio galaxies based on
Ultra Steep Radio Spectrum (USS) and near-IR identification selection
criteria. Also \nobsnoz\ \hzrg-candidates have been observed that did
not show optical emission lines in deep spectroscopic observations at
Keck.

The observations have been carried out over 4 semesters in 4 years
(August 1999 $-$ January 2002) with SCUBA \citep{Holland98spie} at the
James Clerk Maxwell Telescope. We observed at 450\,\micron\ and
850\,\micron\ wavelengths in photometry mode.  The optical depth
$\tau_{850}$ varied between 0.14 and 0.38 with an average value of
0.26.  The data were reduced following standard procedures outlined in
the SCUBA Photometry Cookbook.

\section{Results and Discussion}

\subsection{\hzrgs\ with $z > 3$}
\begin{figure}[t]
\begin{minipage}{8cm}
\psfig{file=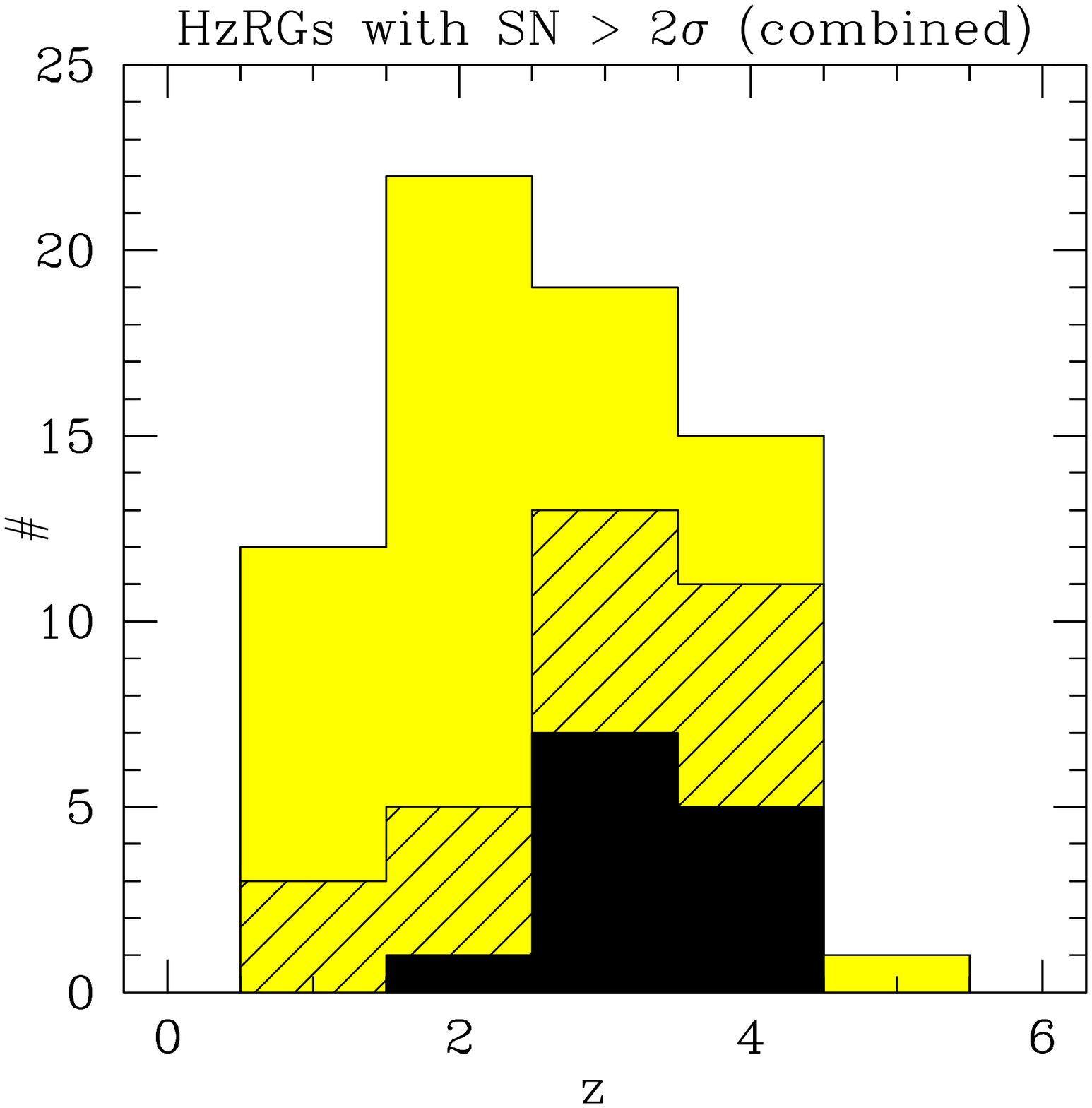,width=6cm}
\end{minipage}
\begin{minipage}{8cm}
\psfig{file=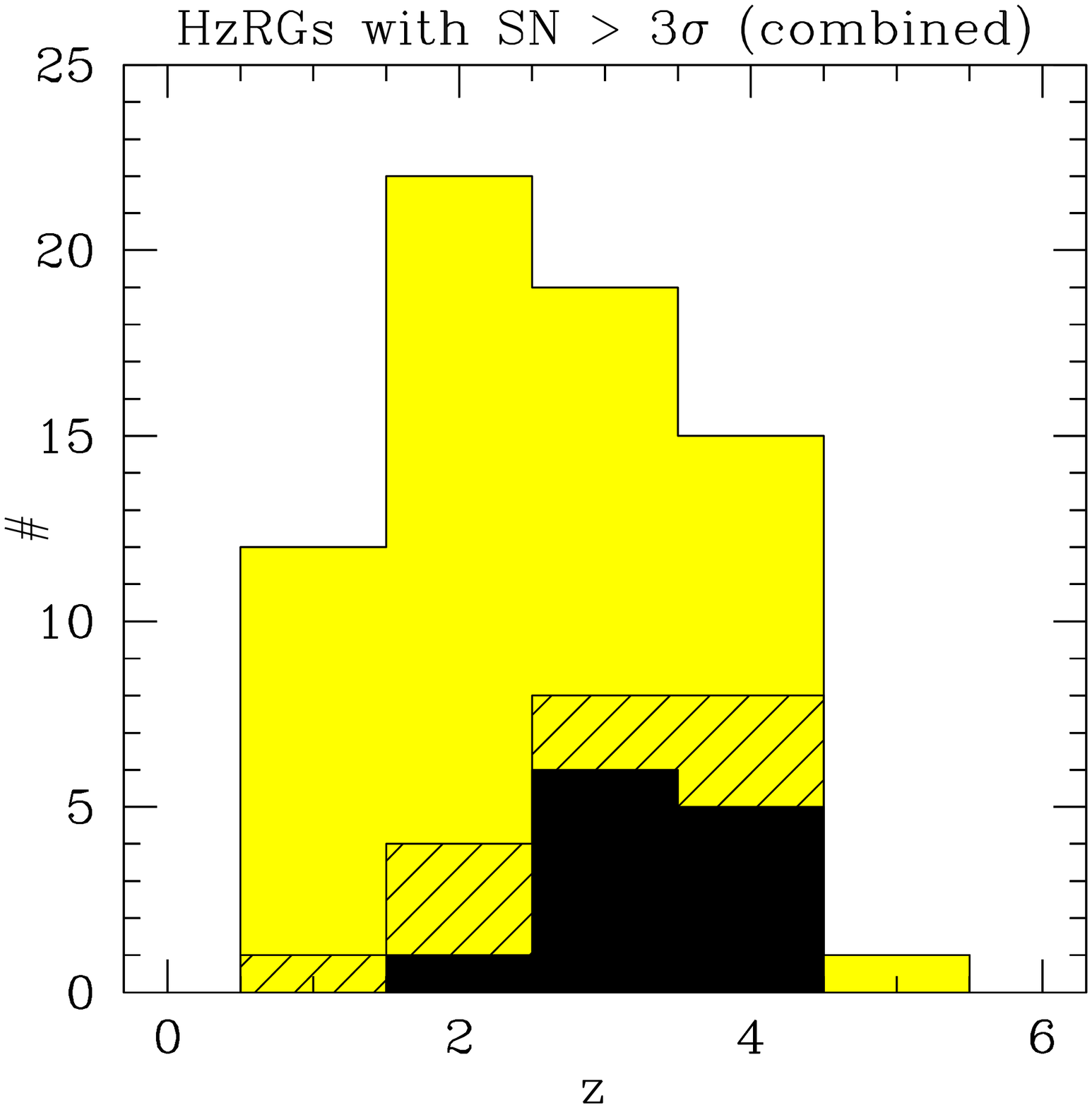,width=6cm}
\end{minipage}
\caption{Histograms of 850\,\micron\ SCUBA detections (dashed) with
S/N $>$ 2 (left) and S/N $>$ 3 (right) versus the total number of
observed radio galaxies (grey) as a function of redshift. The data are
from \cite{Archibald01mnras,Reuland03prepb}, with the detections from
our program shown in black.  At $z > 2.5$ $\sim 50\% - 67\%$ of the
galaxies are detected, as opposed to $\sim 15\%$ at $z < 2.5$.}
\label{dethist}
\end{figure}

In the course of  our program we observed \nobsz\ \hzrgs, of which 
12 were detected with a signal-to-noise ratio larger than 3.  Adding
these data to the sample of \citet{Archibald01mnras} increases the
number of 3$\sigma$-detections at $z > 3$ from 5 to 16.  There was no
attempt to correct for a contribution to the submillimeter continuum
from the non-thermal radio emission, since most sources were selected
using an USS criterion.  Moreover, extrapolation from the radio regime
to submillimeter wavelengths is very uncertain at best, and is likely
to result in an overestimate of the non-thermal flux density due to
steepening of the radio spectrum.  For this reason \citep[and
following][]{Willott02mnras} we have chosen to use the
``un-corrected'' values from \citet{Archibald01mnras}.  Figure
\ref{dethist} shows two histograms representing the number of \hzrgs\
that were observed at each redshift and the number of \hzrgs\ that
were actually detected given a 2$\sigma$ or 3$\sigma$ detection
criterion. Regardless of the actual detection limit used, a marked
increase in the fractional detection rate from $\sim 15\%$ at $z <
2.5$ to $\sim 50\%$ at $ z > 2.5$ is apparent, confirming the result
found by \citet{Archibald01mnras}.

\begin{figure}
\begin{minipage}{8cm}
\psfig{file=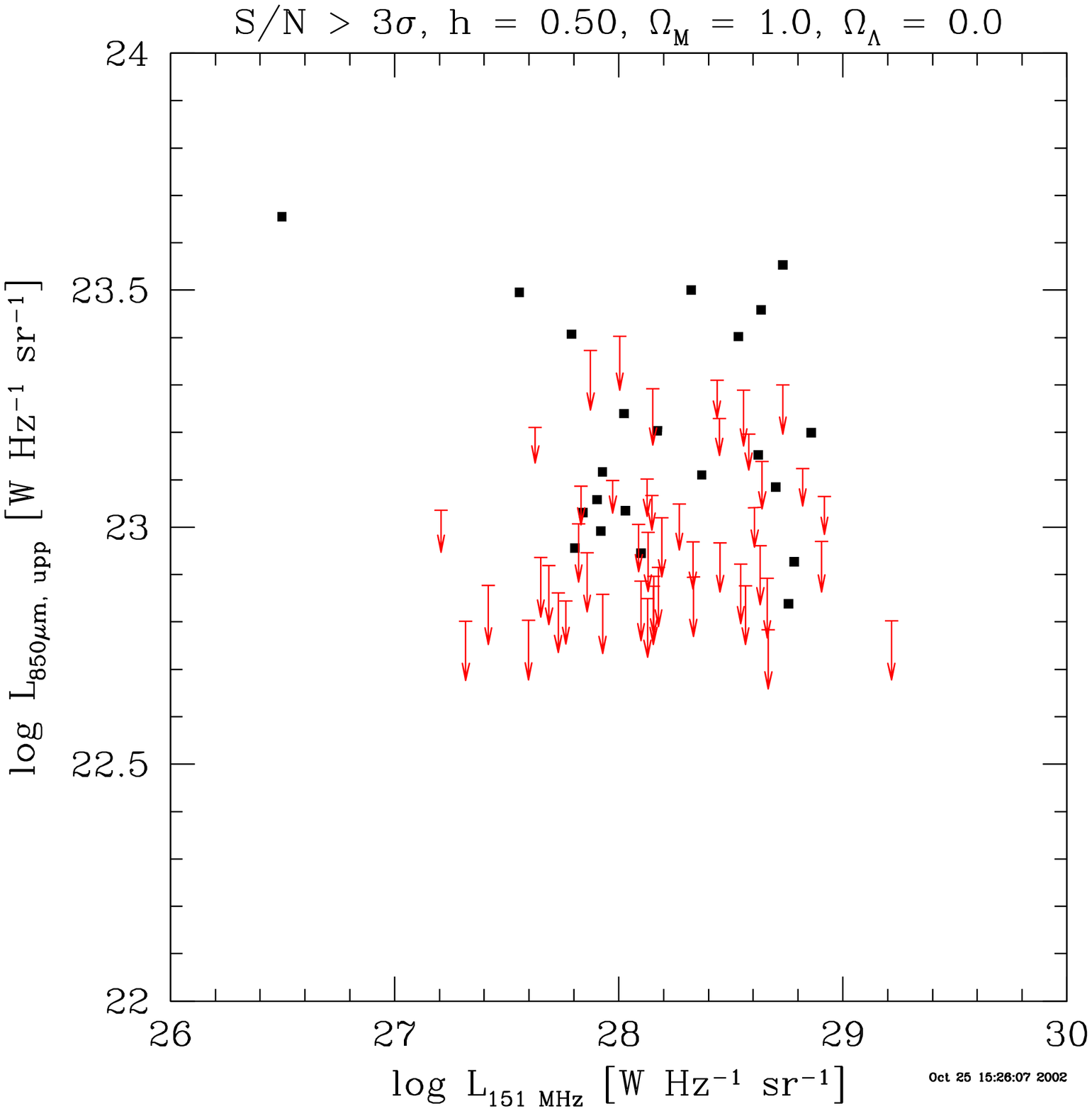,width=6cm}
\end{minipage}
\begin{minipage}{8cm}
\psfig{file=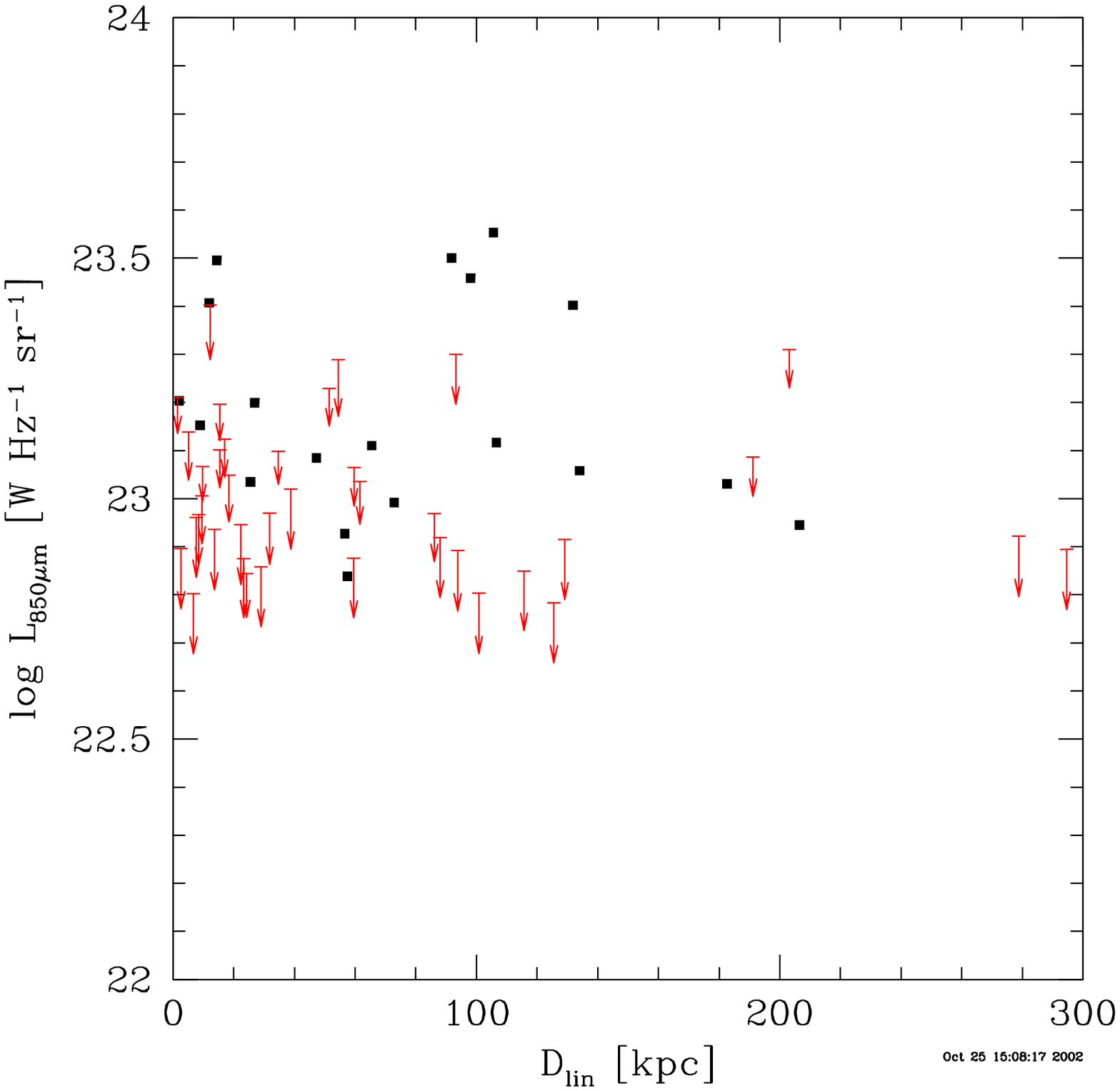,width=6cm}
\end{minipage}
\caption{The left plot shows $L_{850}$ versus radio power. The right
plot shows $L_{850}$ versus linear size. No obvious correlations are
present.  }
\label{L151D}
\end{figure}

We have correlated the inferred submillimeter luminosities with radio
power to search for signs of significant heating by the AGN. No strong
correlation is apparent (Fig. \ref{L151D}). Recently, it was found
that the size of the radio source and $L_{850}$ are correlated for a
sample of radio-loud QSOs at $z = 1.5$, with smaller
\citep[i.e. younger;][]{Blundell99aj} sources being more luminous
\citep{Willott02mnras}.  This effect was attributed to a possible
relation between the jet-triggering event and a short-lived starburst
or quasar-heated dust in QSOs. The effect was not found for a matched
sample of radio galaxies at $z \sim 1.5$.  We have checked whether
there is a correlation between $L_{850}$ and radio source size (age)
in our larger sample. Figure \ref{L151D} shows that they are not
strongly correlated.  However, size of the radio source depends not
only on age but also on jet-power and enviroment which is hard to
correct for in our heterogeneous sample.  Therefore we cannot exclude
that $L_{850}$ and age of the radio source are correlated.

\begin{figure}
\begin{minipage}{4cm}
\hspace{2cm}
\end{minipage}
\begin{minipage}{8cm}
\psfig{file=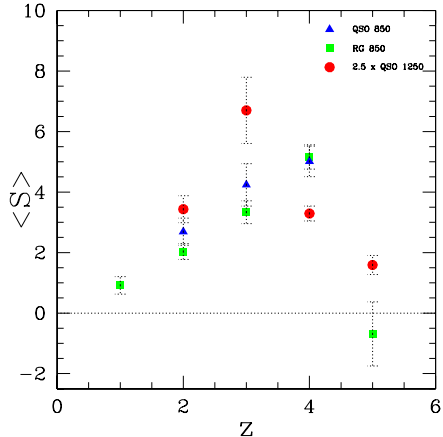,width=6cm}
\end{minipage}
\caption{The average measured (sub-)millimeter fluxdensity binned in
redshift has been plotted for QSOs and radio galaxies.  The squares
represent observations of radio galaxies at 850\,\micron, the
triangles QSOs at 850\,\micron, while the circles represent
observations of QSOs at 1250\,\micron\ multiplied by a factor of 2.5
for easier comparison. }
\label{avgS850}
\end{figure}

The (sub-)millimeter emission of starbursts should be largely indepent of
viewing angle.  Unification schemes \citep[\eg][]{Barthel89apj}
therefore suggest that we can combine observations of \hzrgs\ with
those of QSOs, allowing us to study the evolution of starformation in
their host galaxies over a larger redshift range. Figure \ref{avgS850}
shows the average observed \mbox{(sub-)}millimeter fluxdensity as a
function of redshift for mostly radio-quiet QSO samples
\citep{Carilli01apj,Omont02aa,Isaak02mnras,Priddey02mnras} and \hzrgs.
Reassuringly, the trends mimic each other and they indicate that the
average observed flux density rises to redshifts of $z\sim4$ after
which it declines.

\subsection{Obscured ``no-$z$'' \hzrgs}
During our search for \hzrgs\ among USS sources, we found that 33\% of
the selected canditates were at redshifts $z > 3$
\citep{DeBreuck01aj}. However we also found that $\sim 24\%$, mostly
with a compact radio morphology, do not show clear signs of emission
lines and that a significant fraction ($\sim 10\%$ of the entire
sample) does not even show continuum emission. Possible explanations
for this may be that (i) these are galaxies in the ``redshift desert''
($1.5 < z < 2.3$) where only weak emission lines fall in the optical
window, (ii) they are pulsars, (iii) they are extremely obscured AGN
(iv) or at such high redshift ($z > 7$) that \lya\ is shifted out of
the optical passband.

\begin{figure}
\begin{minipage}{8cm}
\psfig{file=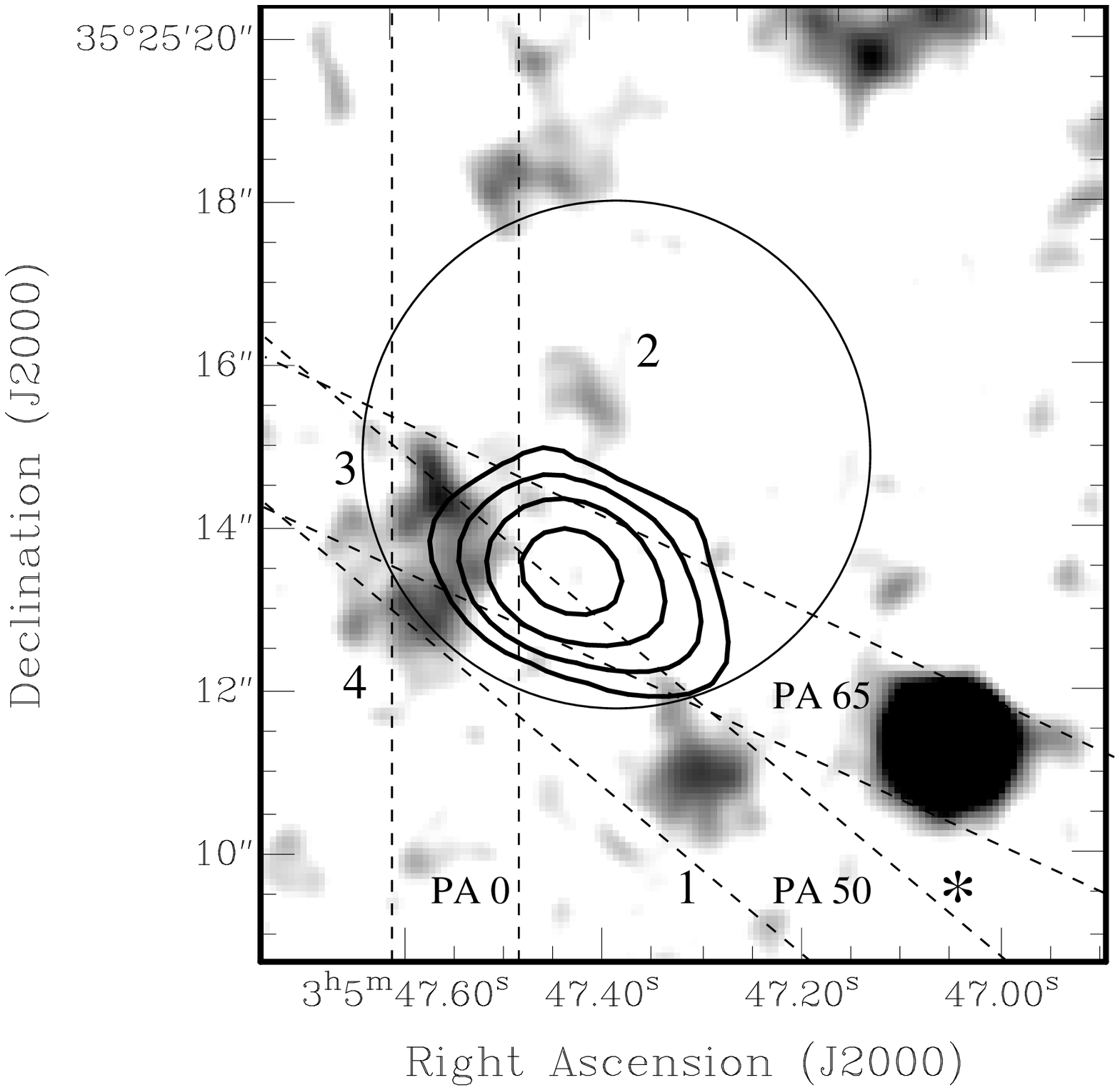,width=6cm}
\end{minipage}
\begin{minipage}{8cm}
\psfig{file=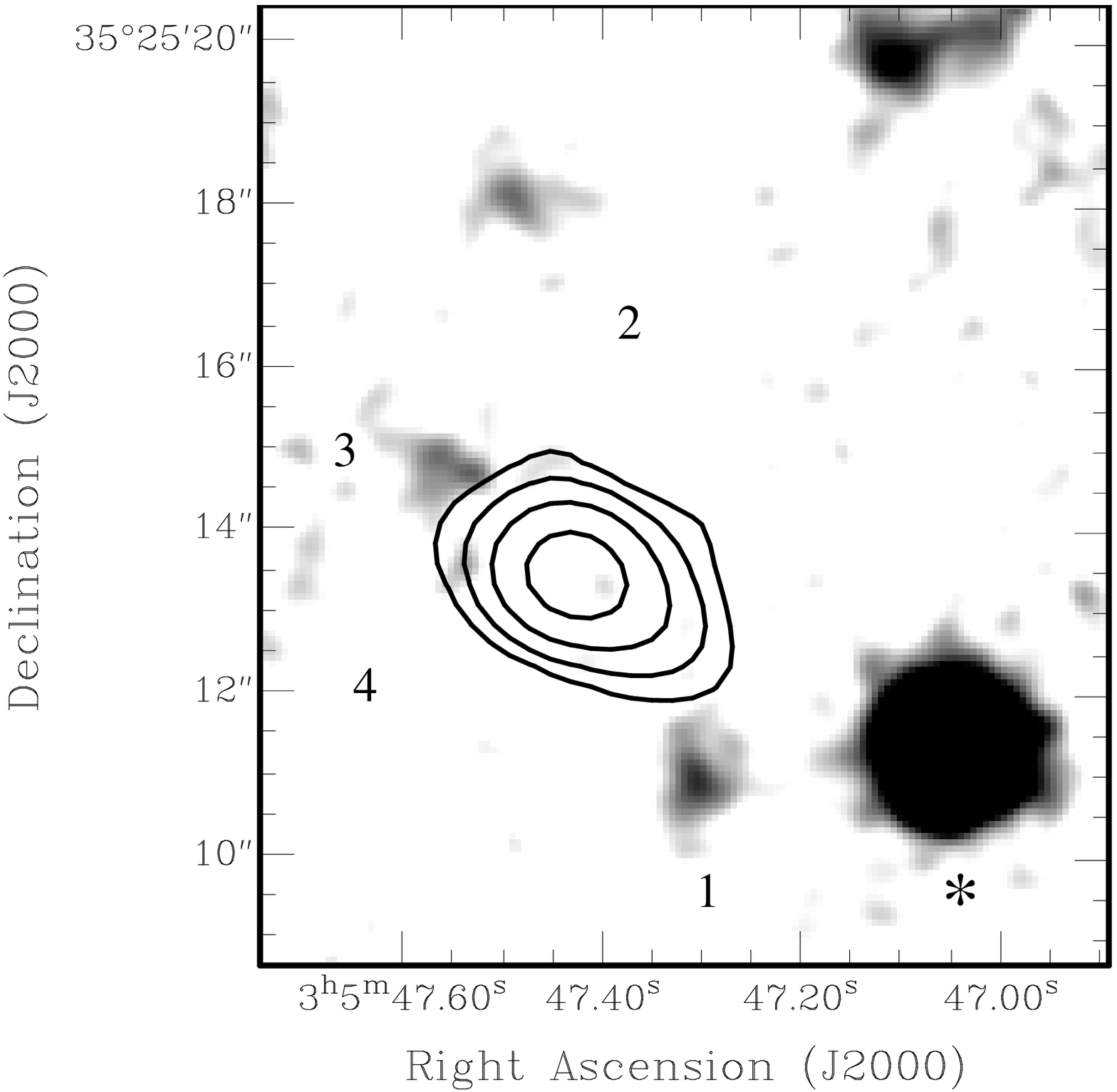,width=6cm}
\end{minipage}
\caption{Keck/NIRC $K$-band (left) and $J$-band (right) images with
4.85\,GHz VLA radio contours overlaid of the \hzrg\ candidate
WN~J0305+3525.  Note the multiple components; objects 2 and 4 are not
detected in the 58\,min $J$-band image. The bright object to the SW is
a spectroscopically confirmed star. The circle with 3\arcsec\ radius
represents the nominal 3$\sigma$ astrometric uncertainty for the
centroid of the 850\,\micron\ emission and dashed lines indicate the
LRIS slit positions.}
\label{0305}
\end{figure}

Figure \ref{0305} shows one of these objects, WN~J0305+3525, which was
studied in detail \citep{Reuland03apj}. It is a strong
\mbox{(sub-)millimeter} detection with $S_{850}$ = $12.5 \pm 1.5$\,mJy and
$S_{1250}$ = $4.2 \pm 0.6$ \,mJy, and appears associated with a number
of faint $K \sim 21$\,mag and $J \ge 23$\,mag objects. The radio
morphology is compact ($\theta < 1.9$\,\arcsec), as is typical of the
other candidates. Based on circumstantial evidence like the near-IR
magnitudes and morphology, the (sub-)millimeter ratios and compact
radio source morphology it was argued that WN~J0305+3525 is a highly
obscured \hzrg\ possibly at a redshift $z = 3 \pm 1$ in a young stage
of its evolution \citep{Reuland03apj}. We have detected 3 more of
these \hzrg-candidates with flux densities of $S_{850} \approx 5 \pm
1$\,mJy resulting in a detection rate of 4 out of 10, very similar to
what is found for the confirmed \hzrgs. This further supports the idea
that they may be young, extremely obscured \hzrgs.

\section{Summary}
\begin{itemize}
\item Submillimeter detection rate appears to be primarily a function
of redshift. If this is interpreted as being due to a change in the
intrinsic far-IR luminosity, it would be consistent with a scenario in
which the bulk of the stellar population of radio galaxies forms
rapidly around redshifts of $z = 3-5$ after which they are more
passively evolving.

\item A significant fraction (10\%) of USS sources may be very
obscured radio galaxies. The obscuration may be related to their young
evolutionary stage.

\item In contrast to the submillimeter sources, \hzrgs\ have
accurately determined redshifts and host identifications.  Even with
the advent of SIRTF this is not likely to change.  Therefore, \hzrgs\
are a key population for studies of galaxy formation in the early
universe, allowing detailed follow-up mm-interferometry observations
to study their dust and gas content.
\end{itemize}

The work of M.R. and W.v.B. was performed under the
auspices of the U.S. Department of Energy, National Nuclear Security
Administration by the University of California, Lawrence Livermore
National Laboratory under contract No. W-7405-Eng-48.

\end{document}